\documentclass[12pt]{article}
\usepackage{epsfig}
\usepackage{axodraw}
\newcommand{\rme}{\ensuremath{\mathrm{e}}}
\def\ltsim{\lower3pt\hbox{$\, \buildrel < \over \sim \, $}}
\def\gtsim{\lower3pt\hbox{$\, \buildrel > \over \sim \, $}}

\begin{document}


\baselineskip=16pt
\begin{titlepage}
\rightline{UG--FT--133/01}
\rightline{CAFPE--3/01}
\rightline{hep-ph/0111047}
\rightline{November 2001}  
\begin{center}

\vspace{0.5cm}

\large {\bf 
Signals from extra dimensions \\[2mm]
decoupled from the compactification scale
}
\vspace*{5mm}
\normalsize

{\bf F. del Aguila \footnote{faguila@ugr.es} and 
J. Santiago \footnote{jsantiag@ugr.es}}

\smallskip 
\medskip 
{\it  Departamento de F\'\i sica Te\'orica y del Cosmos and \\ 
Centro Andaluz de F\'\i sica de Part\'\i culas Elementales}
(CAFPE),\\ 
{\it Universidad de Granada, E-18071 Granada, Spain}
\smallskip  

\vskip0.6in \end{center}
 
\centerline{\large\bf Abstract}
Multilocalization provides a simple way of decoupling 
the mass scale of new physics from the compactification 
scale of extra dimensions. It naturally appears, for example, when
localization of fermion zero modes is used to explain the observed 
fermion spectrum, leaving low energy remnants of the geometrical
origin of the fermion mass hierarchy.
 We study the phenomenology 
of the simplest five dimensional model with order one Yukawa couplings
reproducing the standard fermion masses and mixing angles
and with a light Kaluza-Klein quark $Q_{2/3}$ 
with observable new effects at large colliders.

\noindent
\vfill
\begin{center}
 \\[0.2mm]
{\bf Keywords:} \begin{minipage}[t]{9.5cm}  
Field theories in extra dimensions, multilocalization, 
quark masses and mixings, top quark. 
\end{minipage}
\end{center}
\end{titlepage}

\section{Introduction}

Theories with extra dimensions have received a great deal of attention
during the last
years~\cite{Antoniadis:1990ew,Arkani-Hamed:1998rs,Randall:1999vf}. 
They not only help to explain four dimensional puzzles but predict new
physics at observable scales. Thus, the hierarchy of fermion masses
and mixing angles can be related to the localization of the
corresponding zero modes at different points in the extra
dimensions~\cite{Arkani-Hamed:1999dc}. 
These models reproduce 
the standard four dimensional fermion spectrum with order one
Yukawa couplings, the small mass ratios and mixing angles resulting
from the small overlapping of the corresponding wave functions in the
extra
dimensions~\cite{Arkani-Hamed:1999dc,Mirabelli:1999ks,Grossman:1999ra,
Huber:2001ug}~\footnote{For another approach to the flavour problem in
extra dimensions see~\cite{Hall:2001rz}.}.   
However, the fermion splitting can induce large flavour
changing neutral currents (FCNC) mediated by the Kaluza-Klein (KK) tower of
gauge bosons~\cite{Delgado:1999sv}, and 
then the experimental constraints on rare
processes involving the first two families banish the masses 
of the first excited gauge bosons and the compactification
scale $M_c$ to high energy. 
A possibility of having remnants of the extra dimensions at low energy
is to decouple the mass of the first KK modes from $M_c$. This can be
done multilocalizing the zero mode, this is localizing it at two
different points in the extra dimensions. Indeed,
it has been proven in five dimensional models with
localized gravity that the addition of mass terms modifies the
localization properties of the fields, with the 
multilocalization of the first KK modes resulting in light
four dimensional masses decoupled from the effective compactification
scale~\cite{Kogan:2001wp}. 
This phenomenon,
however, is general and independent of the gravitational
background. As a matter of fact, it is a usual companion of
nontrivial mass terms, and then can naturally appear in 
flavour models in extra dimensions. 

In Section 2 
we study the simplest
five dimensional model with multilocalized fermions. 
We assume a flat  extra
dimension compactified on $\frac{S^1}{Z_2}$
and step function masses, which can
be thought as a limit of more realistic scalar backgrounds. In fact, 
the corresponding four dimensional
effective lagrangian is the same as for more
elaborate theories~\footnote{This also applies to deconstructing
models in four dimensions~\cite{Arkani-Hamed:2001ca}.}. 
In an Appendix we show that in models with
split fermions 
compactified on $\frac{S^1}{Z_2}$ (like, for instance, the one recently
proposed in~\cite{Kaplan:2001ga}) there is also in general
 multilocalization, and then 
massless chiral fermions and light KK modes decoupled from
$M_c$.
We use these results in 
 Section 3 to
construct a definite model with order one Yukawa couplings 
reproducing the standard quark masses and mixing angles
and with an additional 
vector-like quark of charge $\frac{2}{3}$, $Q_{2/3}$, near the
electroweak scale 
and
observable in forthcoming experiments, being directly produced and/or
modifying the top couplings. Finally Section 4 is
devoted to phenomenological implications and conclusions.

\section{Light Kaluza-Klein fermions in flat space}

The appearance of new states
 parametrically lighter than the compactification scale
is a consequence of multilocalization. It has been
recently discussed in great detail in the case of warped
compactifications for particles with spin smaller than 
two~\cite{Kogan:2001wp}  (for the graviton it was
first studied in~\cite{Kogan:1999wc} and further developed
in~\cite{Mouslopoulos:2000er}.) It can be 
also present, however, 
in a flat background.
We review in this Section fermion
multilocalization in the simplest possible context, a five dimensional
model in flat space with the extra dimension compactified on the
orbifold $\frac{S^1}{Z_2}$.
 This 
is a circle of radius $R$ 
with the $Z_2$ identification $y\to
-y$ or an interval $0\leq y \leq \pi R$ with two boundaries, the orbifold
 fixed points.

The action of a spinor reads in  
this space 
(we use the ``mostly minus''
convention for the metric $(+,-,-,-,-)$ and $\gamma^4=\mathrm{i}
\gamma^5$)
\begin{equation}
S=\int \mathrm{d}^4x\;\int_{0}^{\pi R} \mathrm{d}y\;
\bar{\Psi}[\mathrm{i}\gamma^N \partial_N 
-M(y)] \Psi,\quad N=0,\ldots,4,
\end{equation}
where the Dirac mass, which is odd under the action of $Z_2$,
 can be chosen to have a multi-kink structure in order to provide the
 desired multilocalization
\begin{equation}
M(y)=\left\{ \begin{array}{l} 
M,\quad 0\leq y\leq \pi a, \\ 
-M,\quad  \pi a\leq y\leq \pi R, \end{array}\right. 
\label{Dirac:mass}\end{equation} 
with $0\leq a \leq R$. Note that 
the mass $M$ has to
be real by hermiticity but there is no restriction on its sign.  
The kink-antikink shape for the mass
term~\cite{Georgi:2000wb} 
is recovered in the limit $a\to R$.
The KK reduction is performed in the usual way. We 
 split the five dimensional vector-like fermion into its two
 chiralities 
 $\Psi=\Psi_L+\Psi_R$ satisfying $\gamma^5 \Psi_{L,R}=\mp
\Psi_{L,R}$, and expand in KK modes
\begin{equation}
\Psi_{L,R}(x,y)=\frac{1}{\sqrt{\pi R}} \sum_{n=0}^\infty f_n^{L,R}(y)
\Psi^{(n)}_{L,R}(x).
\end{equation}
Substituting them into the action, the decoupling 
of the quadratic terms follows from the relations
\begin{eqnarray}
\int_0^{\pi R} \mathrm{d}y\; \frac{f_n^L f_m^L}{\pi R}&=&
\int_0^{\pi R} \mathrm{d}y\; \frac{f_n^R f_m^R}{\pi R}=\delta_{nm}, \\
( \partial_y- M(y))f^{L}_n=-m_n f_n^{R}&,&
(- \partial_y- M(y))f^{R}_n=-m_n f_n^{L}\label{first:ode}.
\end{eqnarray}
Then the 
corresponding  four dimensional lagrangian describes a chiral zero 
 mode plus a tower of vector-like fermions.
The action of $Z_2$  on $\Psi$ can be chosen to be
 $\Psi(-y)= \gamma^5 \Psi(y)$ or  $- \gamma^5 \Psi(y)$. For
 the sake of concreteness we take the first choice, implying
 that the Right Handed (RH) component 
is even $\Psi_R(-y)=\Psi_R(y)$ and the
 Left Handed (LH) one odd $\Psi_L(-y)=-\Psi_L(y)$. The other
 assignment can be obtained from our results by just interchanging
 $\Psi_L$ and $\Psi_R$ and replacing $M$ by
 $-M$.
 The coupled
first order differential equations (\ref{first:ode}) 
imply the second order one
\begin{equation}
(-\partial_y^2- M^\prime(y)+M^2)f_n^{R}=m_n^2 f^{R}_n,
\label{diffeq:quad}
\end{equation}
where the prime stands for $\partial_y$ and
\begin{equation}
M^\prime (y)= 2 M[\delta(y)-\delta(y-\pi a)+\delta(y-\pi
R)].\label{mass:derivative} 
\end{equation}
$f_n^{L}$, which satisfies the same equation but with $M\to -M$, can
be also obtained 
from $f_n^R$ for $n\neq 0$ 
 using the second equation in (\ref{first:ode}):
\begin{equation}
f_n^{L}=\frac{1}{m_n}( \partial_y+ M(y))f^{R}_n.
\end{equation}

\begin{figure}[ht]
\begin{center}
\epsfig{file=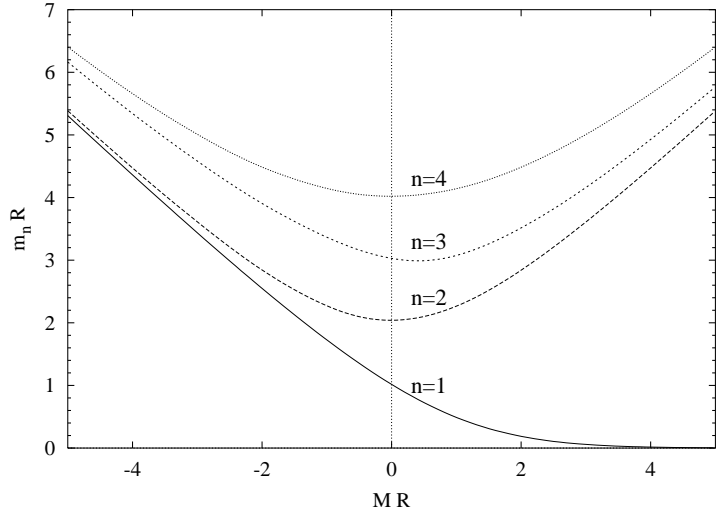,height=7cm}
\end{center}
\caption{Values of the masses of the first KK modes as a function of
the five dimensional mass in units of $1/R$. We have fixed $a=R/2$.
\label{spectrum:fig}}
\end{figure}

The solution of the  Schr\"{o}dinger equation (\ref{diffeq:quad}) 
 for $f_n^R$
is obtained imposing
 the corresponding boundary conditions at $y=0,\pi a$ and $\pi R$,
\begin{eqnarray}
f^{R\;\prime}_n(0)&=&- M f^R_n(0), 
\\
f^R_n(\pi a-\epsilon)&=&f^R_n(\pi a+\epsilon), 
\\
f^{R\;\prime}_n(\pi a+\epsilon)-f^{R\;\prime}_n(\pi a-\epsilon)
&=&2 M f^R_n(\pi a), 
\\
f^{R\;\prime}_n(\pi R)&=& M f^R_n(\pi R), 
\end{eqnarray}
where the limit $\epsilon \to 0$ is understood. 
 The $Z_2$ projection leaves a chiral zero mode with 
 even chirality,
 \textit{i.e.} $f_0^R$. The odd boundary
conditions, which  imply the vanishing of the wave function at the
orbifold fixed points, are not compatible with a massless mode. The zero
mode 
wave function reads
\begin{equation}
f_0^\mathrm{R}(y)=\left\{\begin{array}{l} A \exp[-M (y-\pi a)],\quad 0\leq
y\leq \pi a, \\
A \exp[M(y-\pi a)],\quad \pi a\leq y \leq \pi R,
\end{array}\right. 
\end{equation}
where 
$A=\left[ 
\frac{ 2 M \pi R}{\exp[2M\pi a]+\exp[2 M \pi (R-a)]-
2}
\right]^{1/2}$ is a normalization constant. Thus it is  
exponentially
localized at both orbifold fixed points 
for $M>0$ and at 
$y=\pi a$ (intermediate brane) 
for $M<0$.
(The opposite would happen in the case of an even LH zero mode.)
All the other KK
fermions are vector-like, with the first massive mode $f_1^{L,R}$
having distinctive properties
in the case of 
multilocalization. This occurs if 
the parameters in the potential,
Eqs. (\ref{diffeq:quad},\ref{mass:derivative}), satisfy
\begin{equation}
2 M \pi a(R-a)> R\label{multi:cond}.
\end{equation}
Obviously, the case $a=R$ does
not fulfil Eq. (\ref{multi:cond}) and there is no
 multilocalization for any value of the mass parameter. 
The phenomenology of this case~\cite{Grossman:1999ra,Huber:2001ug,
Kaplan:2001ga,delAguila:2000kb},    
which is also quite interesting, 
can be obtained from our results taking the continuous limit $a\to R$. 
In the case of multilocalization the 
first excited state is also exponentially localized
\begin{equation}
f^R_1(y)=\left\{\begin{array}{l}
B\left(\rme^{k_1 y}+\frac{k_1+ M}{k_1 - M} \rme^{-k_1
y}\right), \quad 0\leq y \leq \pi a, \\
C\left(\rme^{-k_1 (y-\pi R)}+\frac{k_1+ M}{k_1 - M} \rme^{k_1
(y-\pi R)}\right), \quad \pi a \leq y \leq \pi R, 
\end{array}
\right. 
\end{equation}
and
\begin{equation}
f^L_1(y)=\left\{\begin{array}{l}
D\left(\rme^{k_1 y}-\rme^{-k_1
y}\right), \quad 0\leq y \leq \pi a, \\
E\left(\rme^{-k_1 (y-\pi R)}-\rme^{k_1
(y-\pi R)}\right), \quad \pi a \leq y \leq \pi R, \end{array}
\right. 
\end{equation}
with
 $B$ ($D$) and $C$  ($E$)  constants fixed by normalization and 
 continuity conditions.
Its mass $m_1^2=M^2-k^2_1$, with $k_1$ the solution of
 the eigenvalue equation
\begin{equation}
k - M -(k + M) \rme^{-2 k \pi R} + M \left[\rme^{-2k
\pi a}+\rme^{-2 k \pi(R-a)}\right]=0,\label{eigenvalue1:eq}
\end{equation}
can be $\ll M^2$. Eq.~(\ref{eigenvalue1:eq})
 has always only one solution with $0< k_1 < M$. The 
extreme case  $m_1\to 0$ ($m_1 \to M$)
corresponds to $2 M \pi a(R-a)\gg R$ $\left(\;2M \pi a (R-a) \sim
 R\;\right)$. 
The rest of the spectrum consists of oscillating states
 heavier than $M$ and with a
spacing of order $\sim \frac{1}{R}$.  
In Fig.~\ref{spectrum:fig} we show the masses of the first KK fermions
 as function of the five dimensional mass $M$ in units of
 $M_c=\frac{1}{R}$ (we have fixed $a=R/2$ for
 illustration~\footnote{In this case the same model 
can be obtained with a kink-antikink mass
compactifying on $\frac{S^1}{Z_2\times
Z_2^\prime}$~\cite{Barbieri:2000vh}.}  ). As can
 be observed,
there is only one light KK vector-like fermion $f_1$ multilocalized
 for $MR> 2/\pi$, see Eq. (\ref{multi:cond}).
In the absence of FCNC
precision data typically require
$M_c\gtsim 4$ TeV, otherwise this
limit can be as large as $\sim 5000$
TeV~\cite{Delgado:1999sv,Rizzo:1999br}.
This makes a priori the effects of the heavy states small compared
to the contribution of the multilocalized one.

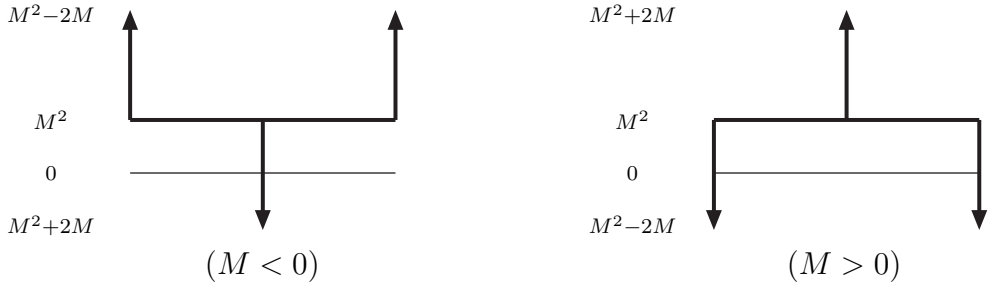
\begin{figure}[ht]
\begin{center}
\begin{picture}(380,120)(-40,-40)
\Line(0,0)(100,0)
\SetWidth{1.5}
\LongArrow(0,20)(0,60)
\LongArrow(50,20)(50,-20)
\LongArrow(100,20)(100,60)
\Line(0,20)(100,20)
\Text(-30,60)[]{${\scriptstyle M^2-2M}$}
\Text(-30,20)[]{${\scriptstyle M^2}$}
\Text(-30,0)[]{${\scriptstyle 0}$}
\Text(-30,-20)[]{${\scriptstyle M^2+2M}$}
\Text(50,-35)[]{$(M<0)$}
\SetWidth{.5}
\Line(220,0)(320,0)
\SetWidth{1.5}
\LongArrow(220,20)(220,-20)
\LongArrow(270,20)(270,60)
\LongArrow(320,20)(320,-20)
\Line(220,20)(320,20)
\Text(190,60)[]{${\scriptstyle M^2+2M}$}
\Text(190,20)[]{${\scriptstyle M^2}$}
\Text(190,0)[]{${\scriptstyle 0}$}
\Text(190,-20)[]{${\scriptstyle M^2-2M}$}
\Text(270,-35)[]{$(M>0)$}
\end{picture}
\end{center}
\caption{Potential $M^2-M^\prime(y)$ 
of the equivalent Schr\"odinger equation for a
 multikink mass term in arbitrary units (for $a=R/2$). 
On the left there is no 
multilocalization ($MR<2/\pi$), in contrast with 
the potential on the right 
which does
multilocalize ($MR>2/\pi$).
\label{ourpotential:fig}}
\end{figure}

\begin{figure}[ht]
\begin{center}
\epsfig{file=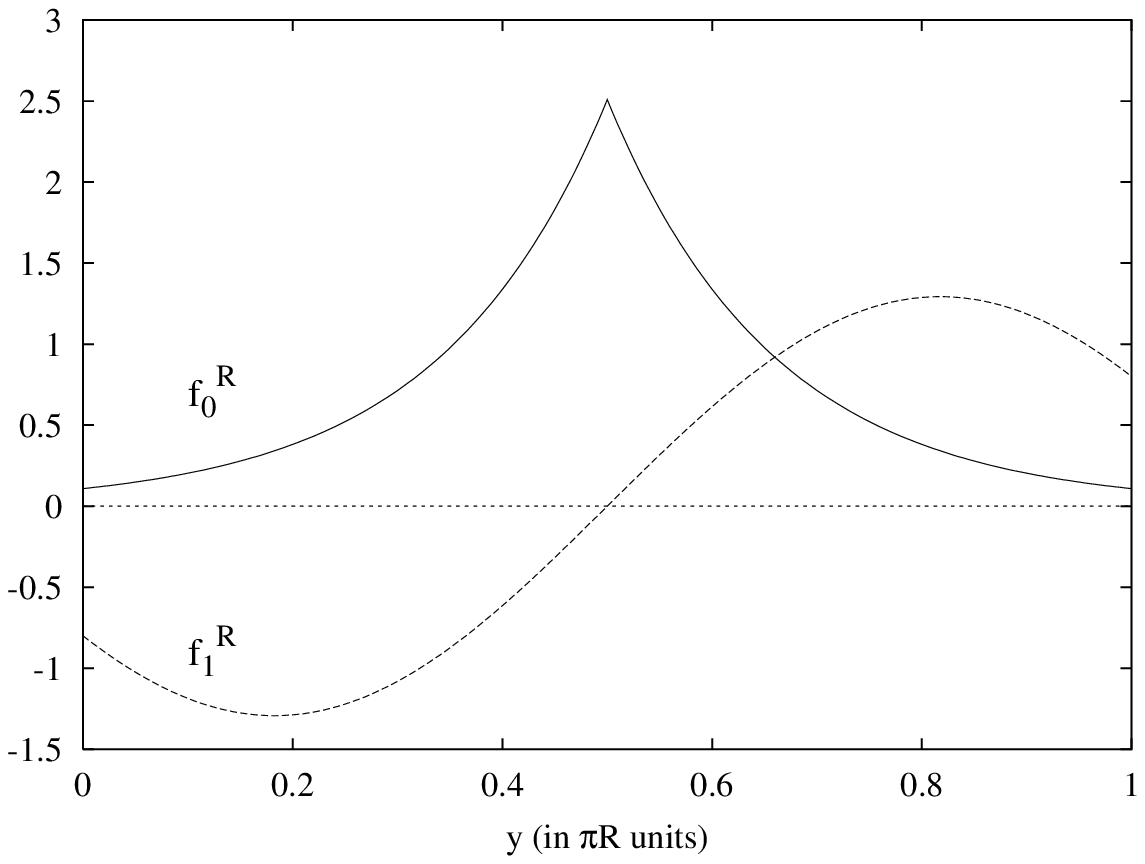,height=5cm}
\epsfig{file=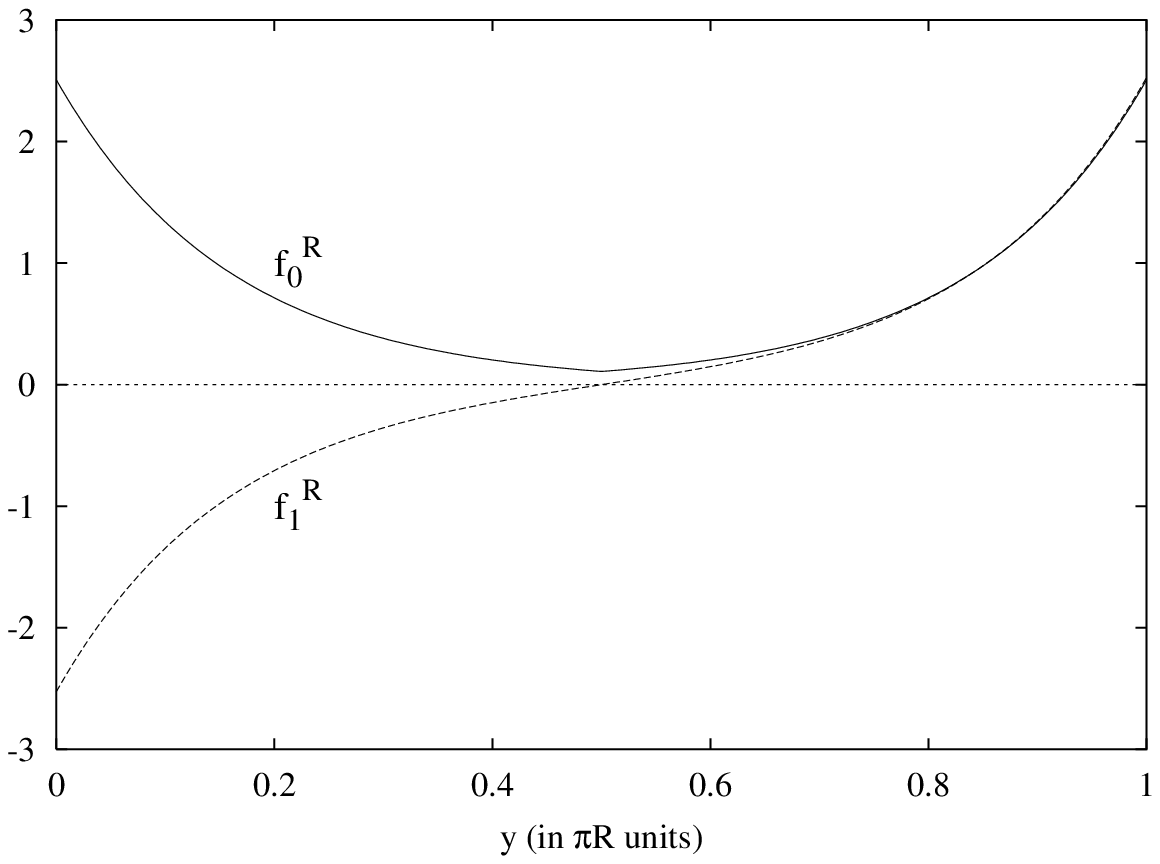,height=5cm}
\end{center}
\caption{Profiles of the massless zero mode $f^R_0$ and the first KK
excitation $f^R_1$ with no multilocalization, $MR=-2$ (left), and
with multilocalization, $MR=2$ (right).
\label{wavefunctions:ours}}
\end{figure}

As has been recently discussed in~\cite{Kogan:2001wp}
the appearance of multilocalization and the related light KK modes can
be easily understood in terms of the shape of the potential in the
corresponding
 Schr\"odinger equation. We review here the argument for completeness.
It has been known for
a long time that domain wall backgrounds create potential wells which
can confine massless fermion zero modes in their
world volume~\cite{Jackiw:1976fn}. In the case that the
potential has a double well shape with each well supporting a bound
state (multidomain wall background), the 
resulting spectrum consists on two  nearly degenerate modes
with a mass splitting proportional to the quantum tunneling
probability between wells. 
In the limit of large separation, 
the lightest state (which is always massless) corresponds to the even
combination of the bound states for the separate wells, while the odd
combination has a mass proportional to the quantum tunneling probability.
The absolute values of the corresponding wave functions only differ in the
intermediate region where they are exponentially suppressed, what
results in the exponentially small mass for the first excited mode. 
In Fig.~\ref{ourpotential:fig} we draw the potential for
the two signs of $M$ in Eq. (\ref{mass:derivative}): 
the case $M<0$ (left) corresponds to the left region in
Fig.~\ref{spectrum:fig} with no multilocalization, and
the case $M>0$ (right) to the right region in the same
Figure with the double well shape of the potential
producing multilocalization for $M$ large enough.
The wave functions of the zero
mode and the first KK excitation are shown in
Fig.~\ref{wavefunctions:ours}. 
Again the left (right) plot corresponds to the left (right) region
in Fig.~\ref{spectrum:fig}. It can be 
observed that in the latter case the difference between the absolute values
of the wave functions are exponentially suppressed. Although we have
discussed the particular case $a=\frac{R}{2}$, multilocalization will
generically occur provided that the scalar background confines the
fermion zero mode at separate points of the extra dimension. This
happens for a wide range of the parameters, location of the
intermediate brane and strength of the potential, as shown in
Eq. (\ref{multi:cond}). 

In our case the
potential consists of delta function wells and barriers because of the
step function mass term we have considered. A possibly more realistic
background would consist of hyperbolic-tangent shaped
masses~\cite{Georgi:2000wb}, with
step functions being considered as a (thin brane) limit. 
From the discussion above it should be clear that in this more
realistic case, provided that the scalar background has a three domain
wall shape, multilocalization can be also present. As an example, we show
in the Appendix that the alternative limit with the intermediate brane
becoming fat and the boundary branes thin (which has been considered
recently in~\cite{Kaplan:2001ga}, where the split fermion idea is
realized on the
orbifold $\frac{S^1}{Z_2}$) 
also presents multilocalization.
Finally we would like to mention the importance that the
orbifold has in our construction. As has been emphasized,
multilocalization will generically occur provided that the scalar
background giving mass to the five dimensional fermions 
has a multidomain wall structure. 
 Orbifold models, which allow to
obtain a chiral spectrum in five dimensions, naturally induce
domain walls at both orbifold fixed points~\cite{Georgi:2000wb}. 
Thus,
any scalar potential with an intermediate domain wall solution in the
orbifold (what can be accomplished by the appropriate boundary
sources~\cite{Kaplan:2001ga})  
automatically presents the multikink structure leading to
multilocalization.


\section{A model of flavour with light Kaluza-Klein quarks}

In this Section we construct a five dimensional model compactified on
$\frac{S^1}{Z_2}$ with the standard fermion content, three
$SU(2)_L\times U(1)_Y$ doublets $q_i$ with LH components even 
and six singlets $u_i,d_i$ with even RH components,
which 
reduces below $M_c$ to the Standard Model
(SM) plus a light vector-like quark of charge $2/3$ with sizeable
mixing with the top quark.
The
 mass terms, $M^{q,u,d}_i$, are 
order the compactification scale with step function shape and 
the Yukawa couplings, $\lambda^{u,d(5)}_{ij}$, order one.
We shall make a detailed numerical discussion to show that there is
 no apparent fine tuning, and to
emphasize its phenomenological relevance. Although the model is
idealized, the four dimensional lagrangian is the same as for more
realistic cases as already stated.

The five dimensional action reads
\begin{eqnarray}
S&=&
\int \mathrm{d}^4x\int_0^{\pi R}\mathrm{d}y\; \left\{ \bar{q}_i
[\mathrm{i}\gamma^N D_N-M^q_i(y)]q_i   +(\mathrm{q\to
u,d})\right.\nonumber \\
&&-\left.\delta(y) [\lambda^{u(5)}_{ij} \bar{q}_i u_j \tilde{\phi}+
\lambda^{d(5)}_{ij} \bar{q}_i d_j \phi+\mathrm{h.c.}]\right\}, 
\end{eqnarray}
where $D_N$ is the covariant derivative and 
to maintain the discussion simple we assume that the Higgs $\phi$ is at
a fixed point, to be concrete at $y=0$. We also assume that all the 
five dimensional
masses $M$ are generated from 
a unique scalar background and therefore that $a$ is
common to all of them.
Performing the KK decomposition and integrating the extra
dimension one obtains the four dimensional lagrangian 
\begin{eqnarray}
\mathcal{L}&=&\sum_{n=0}^\infty \left\{ \bar{q}^{(n)}_i
[\mathrm{i}\gamma^\mu D_\mu-m^{q_i}_n]q^{(n)}_i   
+(\mathrm{q\to
u,d})\right\}\nonumber \\ 
&-&\sum_{n,m=0}^\infty \left\{\lambda^{u(nm)}_{ij} \bar{q}^{(n)}_{Li} 
u^{(m)}_{Rj} 
\tilde{\phi}+
\lambda^{d(nm)}_{ij} \bar{q}^{(n)}_{Li} d^{(m)}_{Rj} \phi+\mathrm{h.c.}
\right\},\label{lag:4}  
\end{eqnarray}
where $D_\mu$ only includes the gauge boson zero modes and $m_{0}=0$.
The effective four dimensional Yukawa couplings are
\begin{equation}
\lambda^{u,d(nm)}_{ij}=\frac{\lambda^{u,d(5)}_{ij}}{\pi R} f^{q_i}_{Ln}
f^{u_j,d_j}_{Rm},\label{yuk:4dim}
\end{equation}
where the wave functions $f$ are evaluated at $y=0$. The 
subscript $L$ $(R)$ for doublets (singlets) is skipped hereafter.
For five dimensional masses
 $M\neq 0$ the exponential localization of the fermion zero modes
can easily give the observed hierarchy of Yukawa 
couplings~\footnote{A similar texture for
the Yukawa matrices is obtained in~\cite{Roberts:2001zy} 
(see also~\cite{King:2001uz} for family symmetries giving the same
mass matrices). The order of the different entries can be slightly
changed varying the five dimensional Yukawa couplings.}
\begin{equation}
\lambda^{u,d(00)}_{ij}
\sim\left (
\begin{array}{ccc}
\epsilon^4 &\epsilon^3 &\epsilon^2 \\
\epsilon^3 &\epsilon^2 &\epsilon \\
\epsilon^2 &\epsilon & 1 
\end{array}
\right) \label{yuk:eps}
\end{equation}
if $f_{0}^{q_i}\sim
f_{0}^{u_i}\sim f_{0}^{d_i}\sim (\epsilon^2,\epsilon,1)$.
This matrix has two zero eigenvalues which 
become order $\epsilon^2$ and $\epsilon^4$  when the order one
 five dimensional Yukawa couplings are also included. 
To obtain such a hierarchy  the third family must be
strongly localized near the Higgs boundary 
and the first two families  at $y=\pi a$, so that they are
suppressed at the fixed point where the Higgs lives. 
As shown in Fig.~\ref{spectrum:fig} for $M$ large
enough (and $0<a<R$) there is also multilocalization and a light KK
vector-like fermion.
In this scenario it seems natural to have $t_R$ more
strongly localized than $t_L$ since the localization of the latter is
the same as 
the one of its doublet counterpart
$b_L$, which has a smaller Yukawa.
This and the fact that with only one light vector-like quark we can account
for large deviations of the SM top couplings make sufficient to
have only $t_R$ multilocalized.

The effective lagrangian for the three light families is obtained
integrating out
 the tower of KK modes in Eq.(\ref{lag:4}). This has been done
in detail in~\cite{delAguila:2000kb}. The corrections 
due to the KK fermions
are proportional to the masses of the corresponding zero modes and to
the inverse of the KK masses squared. Therefore only the top has large
corrections and the main contribution is from the light multilocalized
KK state.
Indeed the largest
top couplings to $Z$ and $W^{\pm}$ in the mass eigenstate basis, 
$X^{L,R}_{tt}$~\footnote{They do not include the electromagnetic
component, being equal to 1 and 0 in the SM, respectively.} and 
$W^{L}_{tb}$, respectively, ($W^{R}_{tb}$ has an extra suppression factor
 $m_b/m_t$) are to first order in 
 $1/m_{n}^2$~\cite{delAguila:2000kb} 
\begin{eqnarray}
X^{L}_{tt}&=&1-
m_t^2 \sum_{k=1}^3|(U^u_R)_{kt}|^2
\sum_{n=1}^\infty 
\left(\frac{f^{u_k}_n}{m^{u_k}_n 
f^{u_k}_0}\right)^2,
\label{xul:flat} \\
X^{R}_{tt}&=&
m_t^2 \sum_{l,k,r=1}^3 V_{tl} (U^q_L)_{lk}^\dagger 
\sum_{n=1}^\infty 
\left(\frac{f^{q_k}_n}{m^{q_k}_n 
f^{q_k}_0}\right)^2
(U^q_L)_{kr}
V_{rt}^\dagger \;,
\label{xur:flat} \\
W^{L}_{tb}&=&V_{tb}
-\frac{1}{2}
m_t^2 \sum_{k=1}^3|(U^u_R)_{kt}|^2 
\sum_{n=1}^\infty 
\left(\frac{f^{u_k}_n}{m^{u_k}_n 
f^{u_k}_0}\right)^2.\label{wl:flat} 
\end{eqnarray}
We neglect corrections suppressed by $\frac{m_i
V_{ib}}{m_t V_{tb}}$ with $i=u,c$, and
the unitary matrices $U$ diagonalize the zero mode mass submatrices
\begin{eqnarray}
(U^q_L)^\dagger_{ik} 
\lambda^{u(00)}_{kl}
\frac{v}{\sqrt{2}}
(U^u_R)_{lj}&=&V^\dagger_{ij} m^u_j, \quad m^u_{1,2,3}=m_{u,c,t},\\
(U^q_L)^\dagger_{ik} 
\lambda^{d(00)}_{kl}
\frac{v}{\sqrt{2}}
(U^u_R)_{lj}&=& m^d_j, \quad m^d_{1,2,3}=m_{d,s,b},
\end{eqnarray}
where $v$ is the
Higgs vev and, in the absence of KK corrections,
$m^{u,d}_j$ are 
 the mass eigenvalues and the unitary matrix $V$ is the 
experimentally measured
CKM matrix.
On the other hand
the corrected top mass reads 
\begin{equation}
m_{t}^\mathrm{phys}=m_t\left(1-
\frac{1}{2} m_t^2 \sum_{k=1}^3 |(U^u_R)_{kt}|^2
\sum_{n=1}^\infty 
\left(\frac{f^{u_k}_n}{m^{u_k}_n 
f^{u_k}_0}\right)^2\right)\label{mt:flat}.
\end{equation}

We are interested in large top mixing and a relatively light exotic
quark consistent with present limits. Both requirements determine $a$
and $M^u_3$ once $R$ is fixed, which is bounded to fulfil the
constraints on FCNC due to the interchange of KK gauge bosons. Thus the
model is essentially fixed if we further assume order one Yukawa
couplings. For example, requiring $W^L_{tb}=0.96$, $m_Q=478$ GeV and
$R=(85 \; \mathrm{TeV})^{-1}$, a 
fit to the data gives
$a=0.51\; R$, 
\begin{eqnarray}
M^{q}_iR&=&(4.01,2.26,-0.30), \nonumber \\
M^{u}_iR&=&(-2.54,0.75,4.77),\label{Mdirac} \\
M^{d}_iR&=&(-3.28,-2.88,-1.88), \nonumber
\end{eqnarray}
 and 
\begin{equation}
\frac{\lambda^{u(5)}_{ij}}{\pi R}=\left(
\begin{array}{ccc}
0.20 & -0.34 & 0.11 \\
-0.10 & 0.06 & -0.08 \\
0.28 & 0.14 & 0.17 
\end{array}\right),
\quad
\frac{\lambda^{d(5)}_{ij}}{\pi R}=\left(
\begin{array}{ccc}
0.35 & 0.10 &  0.46\; 
\mathrm{e}^{-2.77\; \mathrm{i}} \\ 
-0.17 & -0.65 & 0.10 \\
0.25 & 0.13 & 0.25 
\end{array}\right),\label{yuk:5dim}
\end{equation}
where 
the exponential dependence of the quark masses and mixing angles on
the five dimensional masses make these very much constrained,
while a wide range of variation is still 
allowed for the Yukawa couplings.
With these values of the parameters the model reproduces
the observed masses and mixing angles for the known quarks
and predicts a
vector-like quark of charge $2/3$ with 
a mass in the current eigenstate basis  $m^{u_3}_1=462$ GeV.
The other KK fermions have masses $m_n\gtsim 50$ TeV. The observed CP
violation is related to five dimensional Yukawa
couplings which are in general complex, being however enough to take
 $\lambda^{d(5)}_{13}$ complex to obtain the measured CP violation. 
Indeed, integrating
the KK modes~\cite{delAguila:2000kb},
 Eqs.(\ref{xul:flat}-\ref{mt:flat}), we
find (in GeV)
\begin{equation}
m^\mathrm{phys}_{u,c,t}=(
6\times 10^{-3},2.6,165), 
\quad
m^\mathrm{phys}_{d,s,b}=(
11\times 10^{-3},0.26,6.6), \label{masses:our}
\end{equation}
where these masses are to
 be compared with
the running $\overline{\mathrm{MS}}$ masses
 evaluated at the scale $m_t^\mathrm{phys}$. We have
used the experimental values 
in~\cite{Groom:2000in}, taking into account the running up to the
scale $m_t^\mathrm{phys}$~\cite{Barger:1992sf},
\begin{eqnarray}
m_u^\mathrm{phys}=2.9-9.2\;\mathrm{MeV}&,&
m_d^\mathrm{phys}=5.5-16.5\;\mathrm{MeV},\\
m_c^\mathrm{phys}=2.5-2.9\;\mathrm{GeV}&,& 
m_s^\mathrm{phys}=140-310\;\mathrm{MeV},\\
m_t^\mathrm{phys}=161-171\;\mathrm{GeV}&,& 
m_b^\mathrm{phys}=6.2-6.8\;\mathrm{GeV}.
\end{eqnarray}
The corrected CKM matrix writes in the PDG phase
convention~\cite{Groom:2000in} 
\begin{equation}
W^L=\left(
\begin{array}{ccc}
0.9750 & 0.222 & 0.003\; \mathrm{e}^{-1.11 \;\mathrm{i}} \\
-0.222-0.00012 \;\mathrm{e}^{1.11 \;\mathrm{i}} & 0.9742-0.00004 
\;\mathrm{e}^{1.11\;\mathrm{i}} 
& 0.040 \\
0.008 -0.003 \;\mathrm{e}^{1.11 \;\mathrm{i}} & -0.038-0.0007
\;\mathrm{e}^{1.11\;\mathrm{i}}  
& 0.96
\end{array} \right)\label{WL:our}.
\end{equation}
The amount of CP violation can be also given using  
 the Jarlskog
invariant~\cite{Jarlskog:1985ht} 
\begin{equation}
|\mathrm{Im}(W^L_{ub} W^L_{cs} W^{L*}_{us} W^{L*}_{cb})|=2.4\times
10^{-5},
\end{equation}
or the $\beta$ angle~\cite{Groom:2000in} 
\begin{equation}
\sin(2\beta)=\sin\left[2\mathrm{arg}
\left(-\frac{W^{L}_{cd}W^{L*}_{cb}}{W^{L}_{td} 
W^{L*}_{tb}} \right) \right]=0.65,
\end{equation}
which is in agreement with the most recent 
BaBar~\cite{Aubert:2001nu} and
Belle~\cite{Abe:2001xe} measurements
\begin{equation}
\sin(2\beta)=0.59\pm0.14\pm0.05 \quad \mathrm{and} \quad
0.99\pm0.14\pm0.06,
\end{equation} 
respectively.
As can be observed, the masses and mixing angles in
Eqs. (\ref{masses:our},\ref{WL:our}) give the correct experimental
values, only differing appreciably from the minimal SM
\begin{equation}
W_{tb}^L=0.96, 
\end{equation}
and then
\begin{equation}
X^L_{tt}= 0.93. 
\end{equation}
This reflects the lack of unitarity of the CKM matrix 
due to the top mixing with the lightest vector-like KK fermion, which
is predicted to have an observable mass
\begin{equation}
m_Q=478 \;\mathrm{GeV}.
\end{equation}

\begin{figure}[ht]
\begin{center}
\epsfig{file=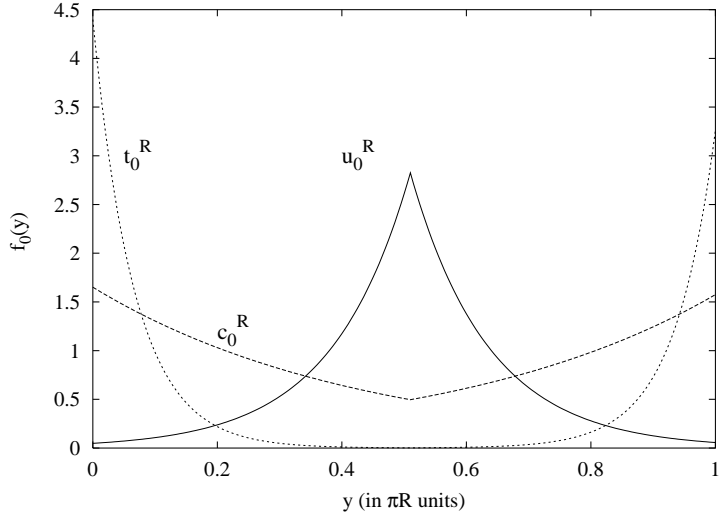,height=7cm}
\end{center}
\caption{Profiles of the RH up quark zero modes for the five
dimensional masses given in the text. Only $t_R$ is
multilocalized.\label{profiles:fig}} 
\end{figure}

\begin{figure}[ht]
\begin{center}
\epsfig{file=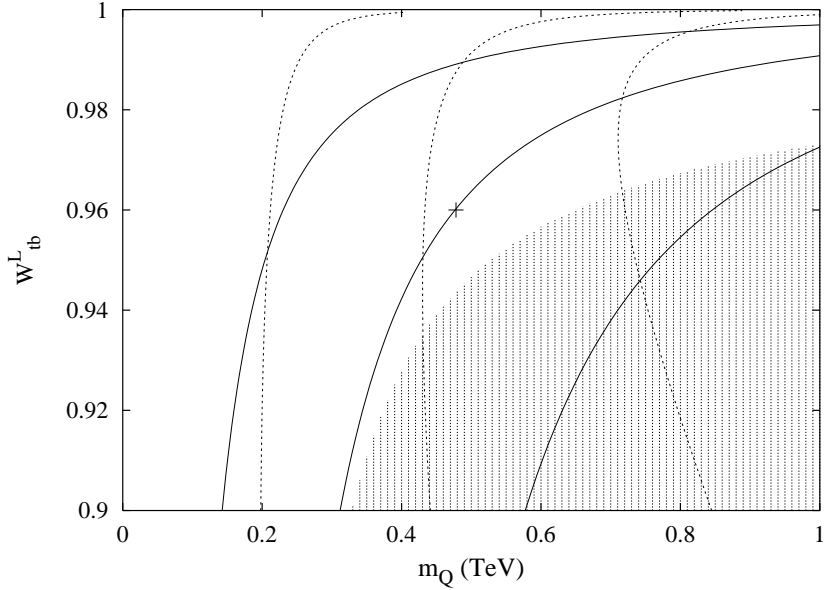,height=8cm}
\end{center}
\caption{Top coupling $W^L_{tb}$ and lightest vector-like quark mass
$m_Q$ as a function of the intermediate brane position $a$ and the
five dimensional mass $M^u_3$. Solid (dotted) 
lines stand for fixed $\frac{a}{R}$ ($M^u_3R$) values, 
from left to right
$0.53,0.51,0.49 $ ($5.55,4.85,4.5$).
We take as in the text
$R=(85\quad \mathrm{TeV})^{-1}$. The shadowed region corresponds to
the 3 standard deviation 
exclusion region for $R_b$.\label{goniometry:fig}}
\end{figure}

Several comments are in order. First, the required minimization to fit
 the experimental values fixing the masses and Yukawa couplings in
Eqs. (\ref{Mdirac},\ref{yuk:5dim}) is essentially determined by the
form of the four dimensional couplings in Eq. (\ref{yuk:4dim}) with their
approximate geometrical form (\ref{yuk:eps}). Second, this
structure is induced by the five dimensional masses through the
 (multi)localization
of the zero modes, see for example Fig.~\ref{profiles:fig}. 
The quarks can be pushed to
 the four dimensional boundary taking the corresponding wave functions
 $f_{1,2,\ldots}\to 0$ and $f_0 \to \sqrt{\pi R}$.
Third, the fit has
been done demanding $W^L_{tb}=0.96$ and $m_Q=478$ GeV. However,
 varying $a$ and $M^u_3$ we can obtain any value of the top mixing and
 the exotic quark mass.
In Fig.~\ref{goniometry:fig} we plot
for arbitrary $W^L_{tb}$ and $m_Q$ the curves with $a$
and $M^u_3$ fixed, respectively, our model depicted by a cross
corresponding to $a=0.51\; R$ and
$M^u_3 R=4.77$. Large departures from the SM in general imply large
 contributions to precision observables. In the absence of
 cancellations a large top mixing with a vector-like quark singlet is
 mainly constrained by $R^\mathrm{exp}_b=0.21664 \pm
 0.00068$~\cite{Groom:2000in}. Using the results of
 Ref.~\cite{Bamert:1996px} we draw
in Fig.~\ref{goniometry:fig} the 3 standard deviation excluded region
for  $R_b$ (the SM prediction being at 1.4 standard deviations from the
 experimental value). In our model $R_b$ is at 2.3 
standard deviations, whereas
the forward-backward asymmetry $A^b_\mathrm{FB}$ prediction is in
slightly better agreement than the SM one and the new contributions
to the oblique parameters S, T, U~\cite{Lavoura:1993np} 
remain within experimental errors~\cite{Groom:2000in}.

\section{Phenomenological implications and conclusions}

We have discussed fermion multilocalization in a five dimensional
model with the flat extra dimension compactified on $\frac{S^1}{Z_2}$
and step function masses. This phenomenon has been studied for particles
up to spin 2 in warped backgrounds~\cite{Kogan:2001wp}, but 
also happens in other backgrounds in the presence of domain wall
masses. Multilocalization allows for a geometrical interpretation of
the hierarchy of fermion masses and mixing angles and at the same time
for the decoupling of vector-like fermions from the compactification
scale, making them observable.

In models with localized fermions at separate points in the extra
dimensions there are a priori large FCNC, requiring compactification
scales up to $M_c\gtsim 5000$ TeV~\cite{Delgado:1999sv}. This bound is
typically reduced for split fermion models with kink-antikink masses  
to $M_c\gtsim 230$
TeV in the case of a boundary Higgs~\cite{Kaplan:2001ga}. 
If these limits are not evaded by the
specific model, and all KK excitations have masses order $M_c$
or higher, no signal of the extra dimensions is expected at
future colliders. However, this is not the case if there is
multilocalization
because the lightest KK
modes decouple from $M_c$. 
The specific model we have worked
out can accommodate 
a new vector-like quark of charge $2/3$ with a mass
$m_Q=478$ GeV and a slightly lower compactification scale $M_c=85$ TeV, 
evading all FCNC constraints. 

 Exotic quarks near the electroweak scale have been extensively
studied in the past~\cite{delAguila:1983fs}. 
 They are present in many
grand unified models, like for instance $E_6$. Models in extra
dimensions with the new vector-like quarks
being the KK excitations of bulk fermions provide 
a natural realization of this possibility.
In general the KK towers of vector-like fermions manifest at low
energy modifying the fermion
mixing~\cite{delAguila:2000aa}, 
with the deviations from the SM
predictions scaling with the masses of the SM quarks
involved~\cite{delAguila:2000kb}. This which may be also indicated by
experiment is naturally realized in these models, in contrast with
former unified scenarios, what together with
 present experimental limits make
the top, and then large colliders, 
the best place to look for these new effects.
In the specific case we have considered 
$Q$ can be directly observed, the reach of Tevatron being
several hundreds GeV and of LHC few
TeV~\cite{delAguila:2000kb,delAguila:1990rq}.  
This vector-like quark also mixes with the top, modifying its 
 couplings $W_{tb}^L=0.96$ and $X^L_{tt}= 0.93$. 
Such a departure of the SM predictions, $W_{tb}^L\ltsim X^L_{tt}=1$, 
cannot be established for $W^L_{tb}$ 
at Tevatron Run II with an accumulated luminosity of
$30\; \mathrm{fb}^{-1}$, the expected accuracy being $7.6
\%$~\cite{Stelzer:1998ni} while it is at the edge for 
 LHC with an expected
precision of $5\%$~\cite{Beneke:2000hk}. Better prospects are
for
$X^L_{tt}$ which will be measured with a precision of $2\%$ at
TESLA~\cite{delAguila:2000fg,Aguilar-Saavedra:2001rg}.  
In Fig.~\ref{goniometry:fig}
we see that this model can reproduce any value of $W^L_{tb}$ and $m_Q$
varying the intermediate brane position and the five dimensional step
function mass for a given compactification scale.
However, banishing cancellations precision data restrict the top mixing
and the exotic quark mass as shown in the Figure.

As a final comment we would like to mention that we have concentrated
on the quark sector, but multilocalization can be
also used for leptons. A first attempt was made
in~\cite{Mouslopoulos:2001uc} (see also~\cite{Huber:2001ug}).

\vspace{1cm}

{\bf Acknowledgments:} This work was supported in part by 
MCYT under contract FPA2000-1558, Junta de Andaluc\'\i a group 
FQM 101 and the European Community's Human Potential Programme under 
contract HPRN-CT-2000-00149 Physics at Colliders. 
J.S. acknowledges MECD for finantial support under an FPU scholarship.

\appendix

\section{Split fermion multilocalization}

In this Appendix we show how multilocalization naturally appears
 in five dimensional models with split fermions and the flat extra
 dimension
compactified on $\frac{S^1}{Z_2}$~\cite{Kaplan:2001ga}.
Fermions are allowed to
 live in the bulk and have a linear odd mass term
\begin{equation}
M(y)=-\frac{M}{R}(y-\pi a),
\end{equation}
with $0\leq a\leq  R$.
A mass term of this kind can be obtained from a bulk scalar with
appropriate
source terms at the orbifold fixed points plus
a step function mass term for $a\neq R/2$~\cite{Kaplan:2001ga}. For 
simplicity, we will consider
the case $a= R/2$. As
before $M$ is constrained to be real by
hermiticity but can have either sign. Let us
consider, for definiteness, that the RH component of the fermion is even and
the LH component odd. Again the opposite parity assignments can be
obtained by just replacing RH by LH and $M$ by $-M$. 
Equation (\ref{diffeq:quad}) is still valid
with the derivative of the mass term being now
\begin{equation}
M^\prime(y)=-\frac{M}{R}+M \pi [\delta(y)+\delta(y-\pi R)].
\end{equation}
The corresponding Schr\"odinger equation in the interior of the
interval is
\begin{equation}
\left[-\partial_y^2+\frac{M}{R}+\frac{M^2}{R^2}(y-\pi R/2)^2 \right]
f_n^R=m_n^2 f_n^R.
\end{equation}
In this case
the potential is 
even under reflections about the
middle point $y=\pi R/2$. Then we can solve 
in the region $\pi R/2<y<\pi R$ and constrain the wave functions to be
alternatively even and odd under this reflection. (Remember that we are
solving for $\Psi_R$ which is even under $Z_2$,  and thus all wave
functions are even at $y=\pi R$.) The boundary conditions read
\begin{equation}
f_{2n}^{R\;\prime}(\pi R/2)=
f_{2n+1}^{R}(\pi R/2)=0,
\quad f_{n}^{R\;\prime}(\pi R)=\frac{\pi M}{2}f_n^R(\pi R),\quad
n=0,1,\ldots.
\end{equation}
In order to solve the Schr\"odinger equation we first make the 
change of variables $t=\sqrt{\frac{M}{R}}(y-\pi R/2)$, which leads to the
equation
\begin{equation}
[-\partial_t^2+(1-\lambda_n+t^2)]f_n^R=0,
\end{equation}
where we have defined the dimensionless variable
$\lambda_n \equiv \frac{m_n^2 R}{M}$. We can now
factorize the asymptotic form $f_n^R(y)= e^{-t^2/2} u_n(y)$,
and make a further change of variables $z= t^2$. The resulting
equation is a confluent hypergeometric equation
\begin{equation}
[z\partial_z^2 +(\frac{1}{2}-z)\partial_z-\frac{1}{4}(2-\lambda_n)]u=0,
\end{equation}
whose general solution can be written in terms of Kummer's functions
$\mathcal{M}(a,b,z)$
\begin{equation}
u_n(z)=A\; \mathcal{M}
\left(\frac{2-\lambda_n}{4},\frac{1}{2},z\right)+B \;z^{1/2}\;
\mathcal{M}\left(1-\frac{\lambda_n}{4},\frac{3}{2},z\right).
\end{equation}
Inverting the changes of variables
 and applying the corresponding boundary conditions we find
\begin{eqnarray}
f_{2n}^R(y)&=&A_{2n}\; e^{-\frac{M}{2R}(y-\pi R/2)^2}
\;
\mathcal{M}\left(\frac{2-\lambda_{2n}}{4},\frac{1}{2},\frac{M}{R}(y-\pi R/2)^2
\right), \\
f_{2n+1}^R(y)&=&A_{2n+1}\; (y+\pi R/2)\; e^{-\frac{M}{2R}(y-\pi R/2)^2}
\;\mathcal{M}
\left(1-\frac{\lambda_{2n+1}}{4},\frac{3}{2},\frac{M}{R}(y-\pi R/2)^2
\right),
\end{eqnarray}
where the masses $m_n^2=\frac{\lambda_n M}{R}$ are obtained solving
the eigenvalue equations for
$\lambda_n$: 
\begin{equation}
\left(1-\frac{\lambda_{2n}}{2}\right)
\mathcal{M}\left(\frac{3}{2}-\frac{\lambda_{2n}}{4},\frac{3}{2},\frac{\pi^2
M R}{4}\right)
-
\mathcal{M}\left(\frac{1}{2}-\frac{\lambda_{2n}}{4},\frac{1}{2},\frac{\pi^2
M R}{4}\right)
=0,
\end{equation}
for the even modes, and
\begin{eqnarray}
\left(1-\frac{\pi^2 M R}{2}\right)
\mathcal{M}\left(1-\frac{\lambda_{2n+1}}{4},\frac{3}{2},\frac{\pi^2
M R}{4}\right)
\quad \quad \quad \quad && \nonumber \\
+\quad \frac{\pi^2 MR}{3} \left(1-\frac{\lambda_{2n+1}}{4}\right) 
\mathcal{M}\left(2-\frac{\lambda_{2n+1}}{4},\frac{5}{2},\frac{\pi^2
M R}{4}\right)
&=&0,
\end{eqnarray}
for the odd ones.

\begin{figure}[!b]
\begin{center}
\epsfig{file=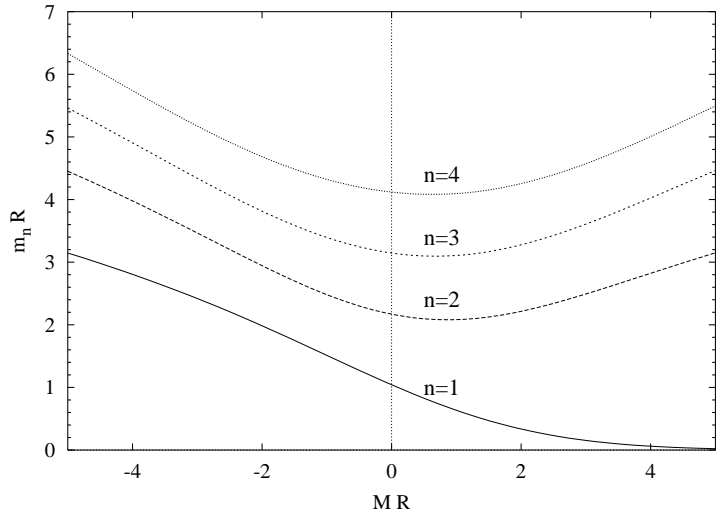,height=7cm}
\end{center}
\caption{Values of the masses of the first KK modes for split fermions
$f_n$ as a function of
the slope of the linear five dimensional mass.
\label{spectrum:kaplan}}
\end{figure}

\begin{figure}[ht]
\begin{center}
\epsfig{file=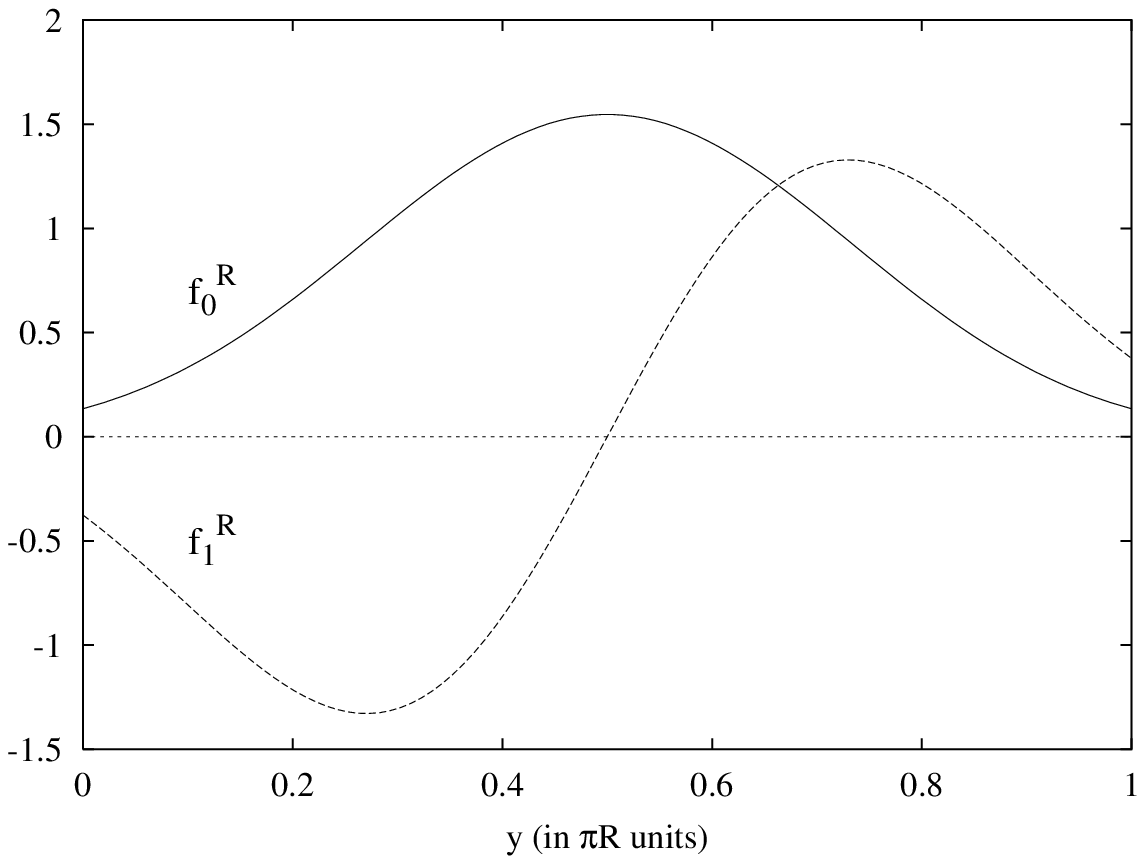,height=5cm}
\epsfig{file=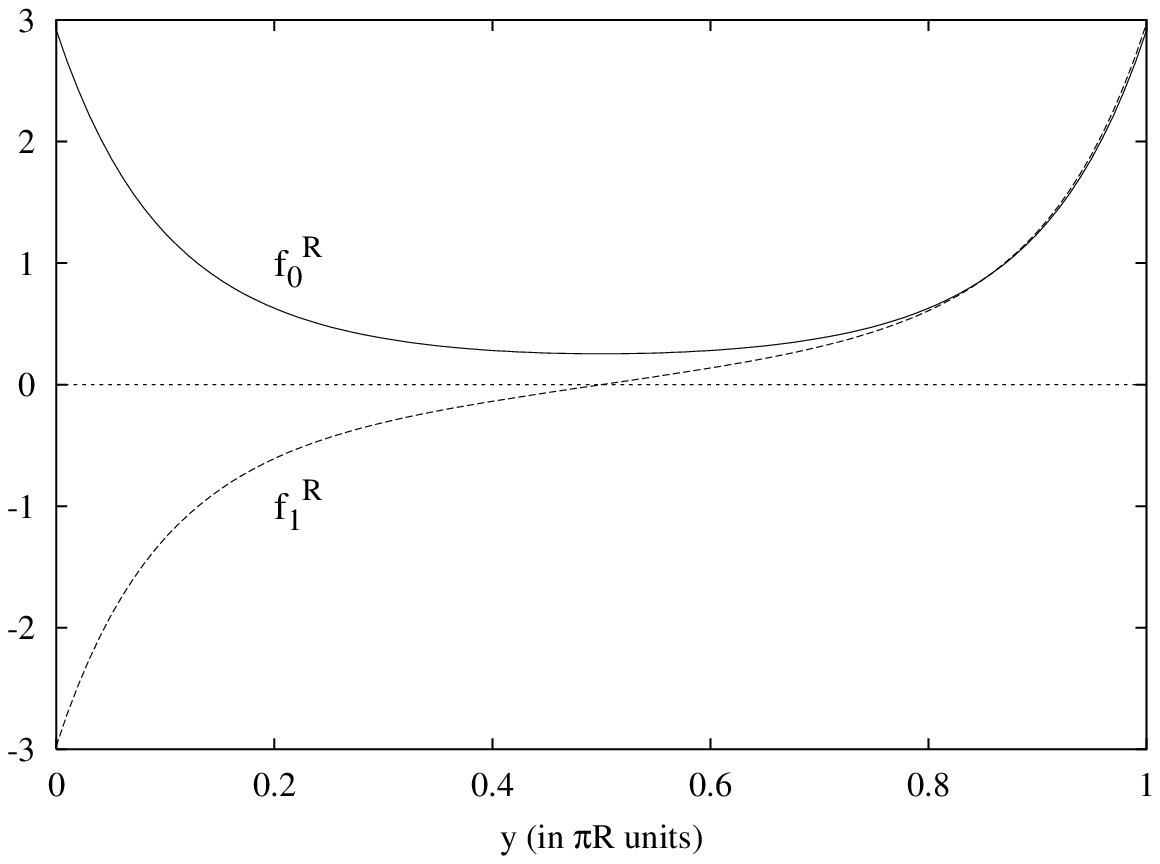,height=5cm}
\end{center}
\caption{Profiles of the massless zero mode and the first KK
excitation for the case of no multilocalization, $MR=-2$ (left),
and multilocalization,
$MR=2$ (right).
\label{wavefunctions:kaplan}}
\end{figure}

To gain some intuition we start considering the massless
zero mode whose wave function can be written in terms of elementary
functions. Using the property $\mathcal{M}(a,a,z)=e^z$ we find that
$\lambda_0=0$ is a solution of the eigenvalue equation for the even
modes. The corresponding zero mode has a gaussian profile
\begin{equation}
f_0^R(y)=A_{0}\; e^{\frac{M}{2R}(y-\pi R/2)^2},
\end{equation}
being localized at $y=\pi R/2$ for $M<0$ and at both fixed points
with exponential suppression at the midpoint if $M>0$. In the latter
case the zero mode is multilocalized and we expect that the first KK
mode be anomalously
light. In Fig.~\ref{spectrum:kaplan} we 
plot the masses of the first few
KK modes as a function of the slope of the five dimensional mass. 
The phenomenon of multilocalization and the
corresponding light KK excitation is apparent for the appropriate $M$
values. We also show in Fig.~\ref{wavefunctions:kaplan} the wave
function profiles for the zero and first modes for the two $M$ signs:
$MR=-2$ with no multilocalization on the left;
and $MR=2$ with multilocalization on the right. There is a close 
resemblance of
this spectrum with the one found in Section 2 
for the case of a multikink mass term. This could
have been anticipated from the shape of the potential with also two
attractive delta functions at the orbifold fixed points.

As a final remark we would like to comment that in the flavour models with
split fermions proposed in the literature 
the appropriate signs 
for the mass terms have been
chosen 
(positive for even LH fields and
negative for even RH ones) 
so as to have gaussian localization around one single point 
and not multilocalization. However, we find no reason for a generic
choice of sign excluding the possibility of multilocalization.

\providecommand{\href}[2]{#2}


\end{document}